\documentclass[superscriptaddress,a4paper,twocolumn,preprintnumbers,showpacs,amsmath,amssymb,prl]{revtex4-1}

\usepackage{graphicx}

\usepackage{color}

\begin{document}
\title{The concept of entropic rectifier facing experiments}

\author{D. Lairez}
\email{lairez@cea.fr}
\affiliation{Laboratoire L\'eon Brillouin, CNRS, CEA, Universit\'e Paris-Saclay, 91191 Gif-sur-Yvette cedex, France}

\author{M.-C. Clochard}
\affiliation{Laboratoire des Solides Irradi\'es, \'Ecole polytechnique, CNRS, CEA, Universit\'e Paris-Saclay, 91128 Palaiseau cedex, France}

\author{J.-E. Wegrowe}
\affiliation{Laboratoire des Solides Irradi\'es, \'Ecole polytechnique, CNRS, CEA, Universit\'e Paris-Saclay, 91128 Palaiseau cedex, France}

\date{\today}
\preprint{preprint}

\begin{abstract}
The transport of molecules in confined media is subject to entropic barriers. So theoretically, asymmetry of the confinement length may lead to molecular ratchets with entropy as the only driving force for the biased transport. 
We address experimentally this question by performing alternative ionic current measurements on electrolytes confined in neutral conical nanopores. In case anions and cations widely differ in size, we show that rectification of ionic current can be obtained that depends on ions size and cycle frequency, consistently with the entropic ratchet mechanism.
\end{abstract}

\maketitle

In fluids, due to random motion of molecules, their directed transport is ordinarily achieved by the mean of an external force field. Brownian motors and ratchets\,\cite{Astumian:1997aa, Reimann:2002aa, Hanggi:2009kx} fulfill this task without net bias, not only despite random motion but thanks to it.
Initially viewed as some thermodynamical curiosities for physicists\,\cite{Bridgman:1928aa, Brillouin:1950aa, Feynmann_Ratchet}, these brownian machines have gained interest when they proven to be relevant in many biological systems, from muscle contraction\,\cite{Vale:1990aa,Astumian:1994aa} to ionic pumps\,\cite{Tsong:2002aa}, including voltage-gated channels\,\cite{Fadda:2013fk}. Nowadays, advances in nanodevices fabrication stimulate bio-inspired realizations mostly concerning smart ions transport and pumping that likely promises the more challenging applications (e.g. keep in mind water desalination issues). 

Ion pumps typically work with nanoporous membranes submitted to a periodic voltage, which is unbiased in average but allows the machines to fulfill the second law of thermodynamics\,\cite{Bridgman:1928aa, Brillouin:1950aa, Feynmann_Ratchet}. Then, these machines are mostly founded on random motion of ions in a sawtooth landscape of potential that materializes the ratchet. All efforts to produce bio-inspired ionic current rectifiers lie in this materialization\,\cite{Hou:2012aa}.
Until now, concrete realizations\,\cite{Siwy:2002fk, Siwy:2006,Cervera:2006,sheng2014,sheng2014,Zhang:2015aa,Madrid:2015aa,Jiang:2016aa} generate the ratchet by means of electrostatic charges on pore-wall, the asymmetry arises from a charge gradient or from the conical shape of nanopores that is felt by ions through electrostatic interactions\,\cite{Kosinska:2008aa}.

In this letter, we consider experimentally a more direct and universal way for any particles to feel the shape of nanopores that could come from entropic barriers\,\cite{Reguera:2006aa}. The idea is the following. For ions with diffusion coefficient $D$, in solution at concentration $c$, moving in a potential $U$, the current $J$ is the sum of brownian diffusion $-D\nabla c$ and drift motion $-Dc\nabla U/kT$: 
\begin{equation}\label{eq1}
J=-D(\nabla c+ c\nabla U/kT)
\end{equation}
In confined systems, entropy changes are crucial and the thermodynamic potential to consider in Eq.\,\ref{eq1} is more suitably the free energy $F=U-TS$.
The substitution of $U$ by $F$ introduces entropic forces, which are palpable and have proven to be of great importance in nanoscopic systems (see e.g. the elasticity of polymer chains\,\cite{deGennes:1996} or the depletion force in colloids\,\cite{Mao:1995aa}).
In a narrow channel, if the transverse equilibrium is rapidly reached, the problem can be treated in 1D using the longitudinal probability density $p(x)= c(x) A(x)$, with $A(x)$ the cross-section area. In ideal solutions, the number of microstates for one given ion is $\Omega \propto A(x)/a$, with $a$ the area unit. The corresponding entropy is $k\ln(A(x)/a)$ and Eq.\,\ref{eq1} without external potential leads to the Fick-Jacobs equation\,\cite{Zwanzig:1992aa}:
\begin{equation}\label{FJeq}
J_x=-D\left({p'(x)-p(x)\frac{d\ln(A(x)/a)}{dx}}\right)
\end{equation}

\begin{figure}[!htbp]
\centering
\includegraphics[width=0.98\linewidth]{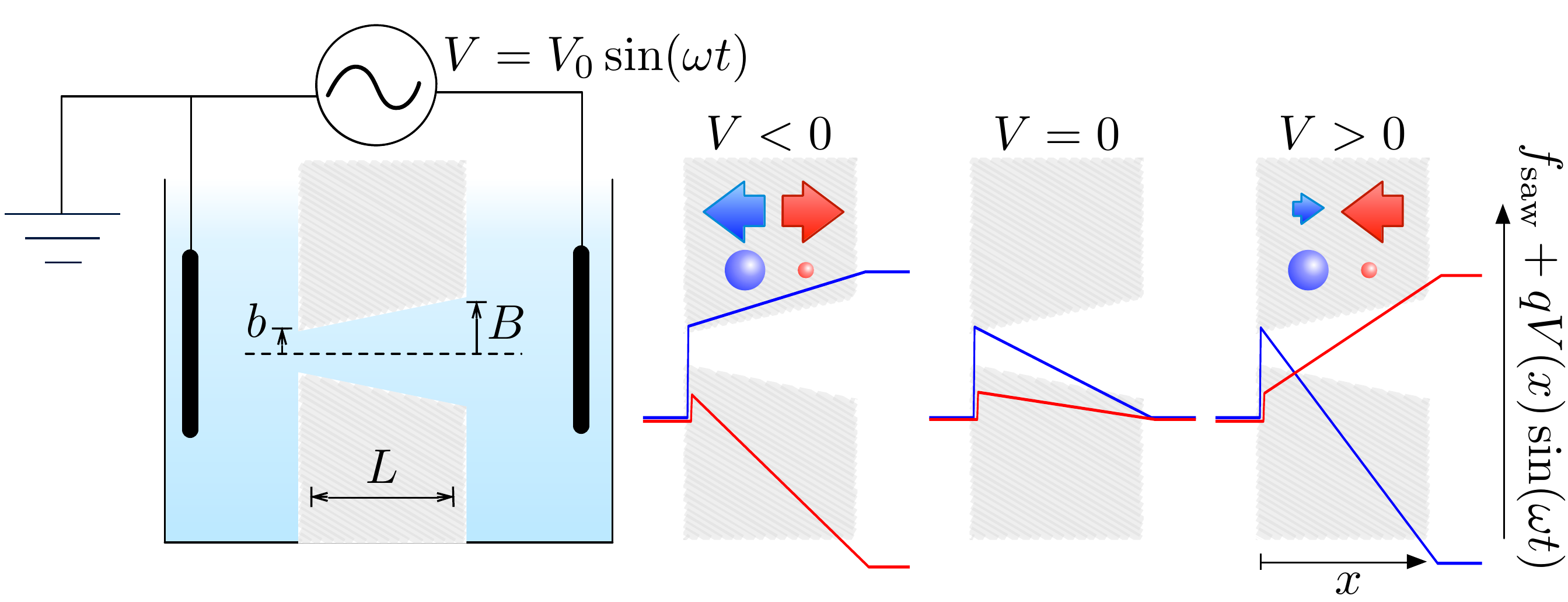}
\caption{Conical nanopore as entropic rectifier: without transmembrane voltage, the sawtooth free energy potential $f_{\text{saw}}$ is repulsive for both large anions (blue) and small cations (red), but smaller for the latter. For $V>0$, the energy barrier for anions at the pore entrance (tip of cone) is higher than for cations when $V<0$. One gets an ionic current diode. In case anions and cations have the same size, the entropic barriers have the same height and no rectification is expected.}
\label{fig1}
\end{figure}

Eq.\,\ref{FJeq} is still insufficient to rectify ionic current because charge carriers are dual and the entropic contribution to the free energy is repulsive for both species (contrary to the potential due to pore-wall charge). Thus when transport is facilitated for anions it is impeded for cations and vice versa. The missing piece is related to the depletion layer\,\cite{Mao:1995aa}: the center of mass of ions cannot approach the wall to distance smaller than their radius $r$, so the accessible area is rather $\approx(\sqrt{A(x)}-r)^2/a$\,\cite{Rubi:2010aa, Reguera:2012aa}. Then, if the pore section is sawtooth, one gets size-dependent entropic barriers that can be used to split ions regarding to their size\,\cite{Reguera:2012aa}. Also, if this idea is correct, thus it should be possible to rectify the ionic current passing through geometrically asymmetric nanopores in case anions and cations widely differ in size (Fig.\,\ref{fig1}). This is what we experimentally address and report in this letter. Our experiments consist in ionic current measurements as a function of the applied sinusoidal voltage (cyclic voltammetry) through conical nanopores similar to those used in ref.\,\cite{Siwy:2002fk, Siwy:2006} using polyacids as charge carrier. We show that at pH conditions insuring the electric neutrality of pore-wall, a clear ionic current rectification can be still obtained that depends on ion size and cycle frequency.

\emph{Experimental details: } Membranes with multiples conical nanopores are prepared similarly to ref.\,\cite{Tasserit:2010fk} by heavy-ion irradiation (Kr$^{28+}$, 10.4\,MeV, fluence 10$^7$\,cm$^{-2}$, performed at GANIL, France) of $L=6$\,$\mu$m thick polyimide foils (Kapton HN) and single-side chemical etching of ion-tracks. The etching was performed at $50^\circ$C using a two chambers conductivity cell, one filled with NaOCl-15\% etching solution, while the other with 1\,M KI neutralizing solution. Transmembrane voltage of +1\,V (with respect to the grounded neutralizing compartment) is applied to detect the breakthrough and help the neutralization of reactant as soon as the opening is achieved.
Doing so, conical nanopores are obtained with a narrow size distribution, small opening radius $b\simeq 2$\,nm and large opening radius $B\simeq 100$\,nm. 
Ionic current were measured using a Biologic-SP200 potentiostat with two Ag/AgCl electrodes. After etching, the two compartments of the conductivity cell were filled with the same electrolyte solution. In order to ensure that the electrode voltage corresponds to the actual transmembrane voltage, the area of the membrane was reduced down to 12.6 mm$^2$ (i.e. $1.26\times10^6$ pores) allowing its impedance to be larger than 95\% of the total (see in Fig.\,\ref{fig2} the current drop due to the membrane). Cyclic voltammetry was performed by applying a $\pm0.5$\,V sinusoidal voltage (small opening of pores at the grounded compartment) and measuring the current. Cycling was repeated until a stationary behavior  (typically 2 or 3 periods). Only these last cycles are shown and discussed in the following. Reversibility was checked after each sample by measuring the membrane behavior with a reference electrolyte (acetic acid 0.1\,M).

\begin{figure}[!htbp]
\centering
\includegraphics[width=0.48\linewidth]{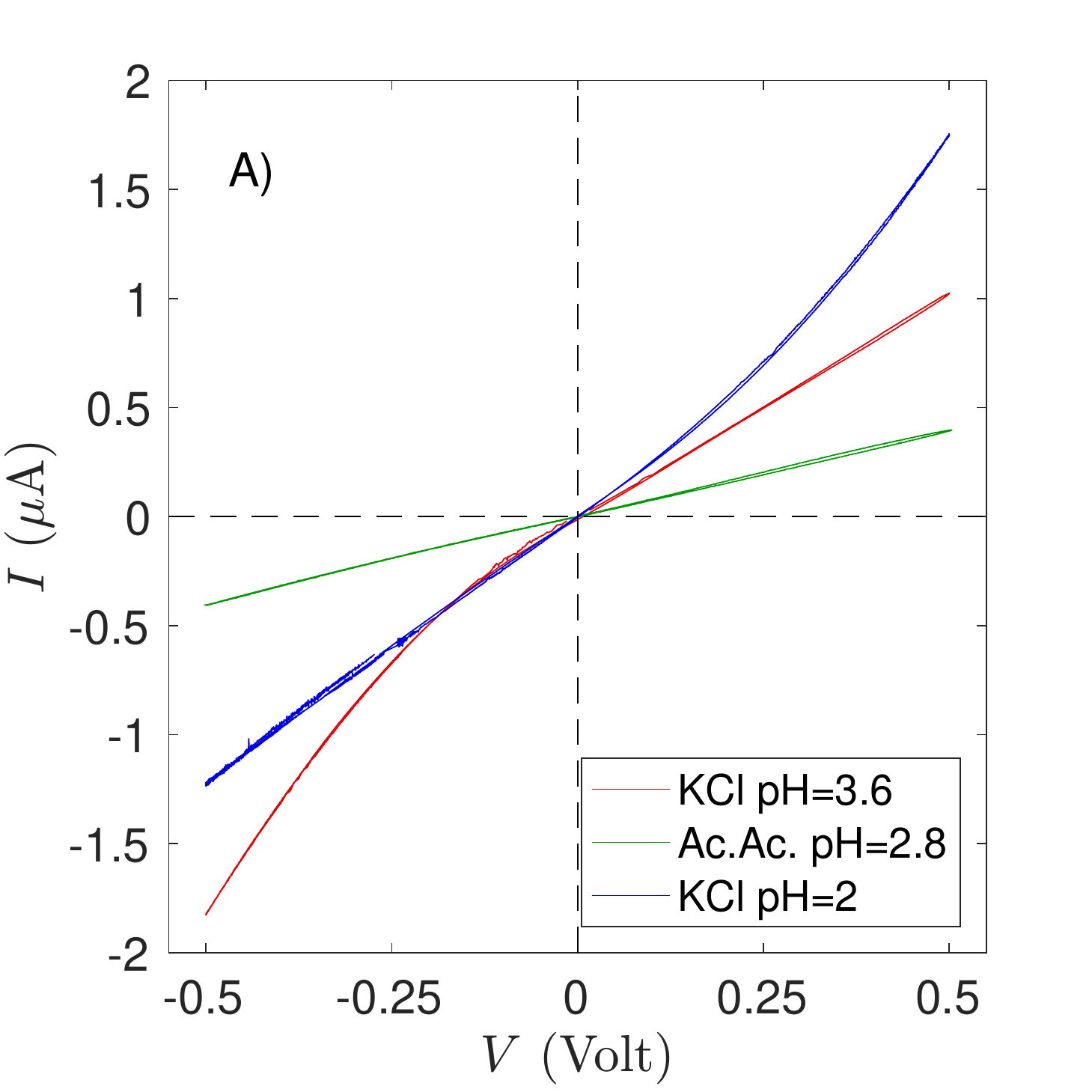}
\includegraphics[width=0.48\linewidth]{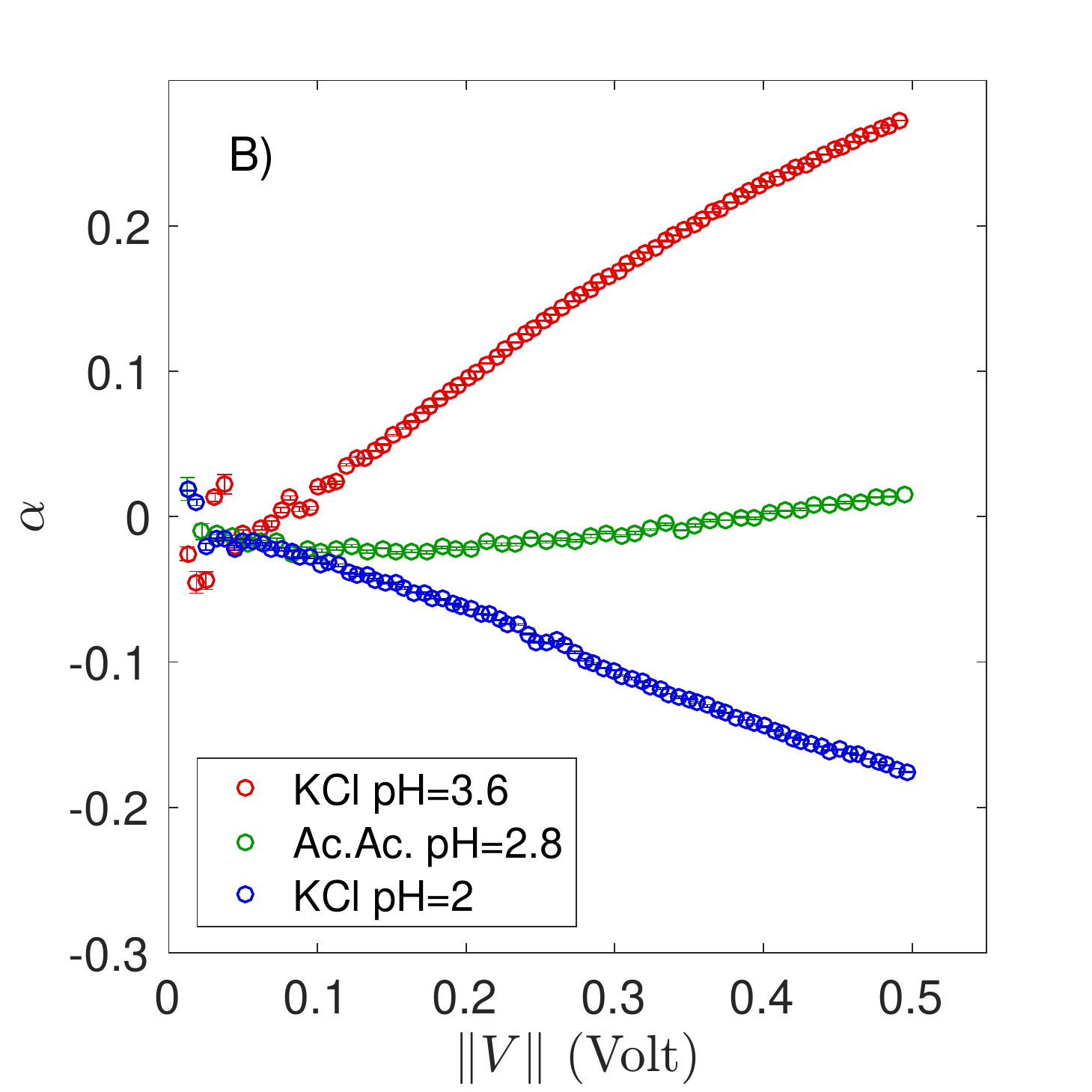}
\caption{Cyclic voltametry for conical nanopores filled with KCl 0.1\,M solutions at pH=3.6 and 2, respectively and acetic acid solution (c=0.1\,M, pH=2.8). A) Current $I$ vs. voltage $V$ measured at 1/32\,Hz; B) Rectification efficiency $\alpha$ vs. absolute value of voltage $\|V\|$ (Eq.\,\ref{alpha}).}
\label{figpH1}
\end{figure}

\begin{figure}[!htbp]
\centering
\includegraphics[width=0.48\linewidth]{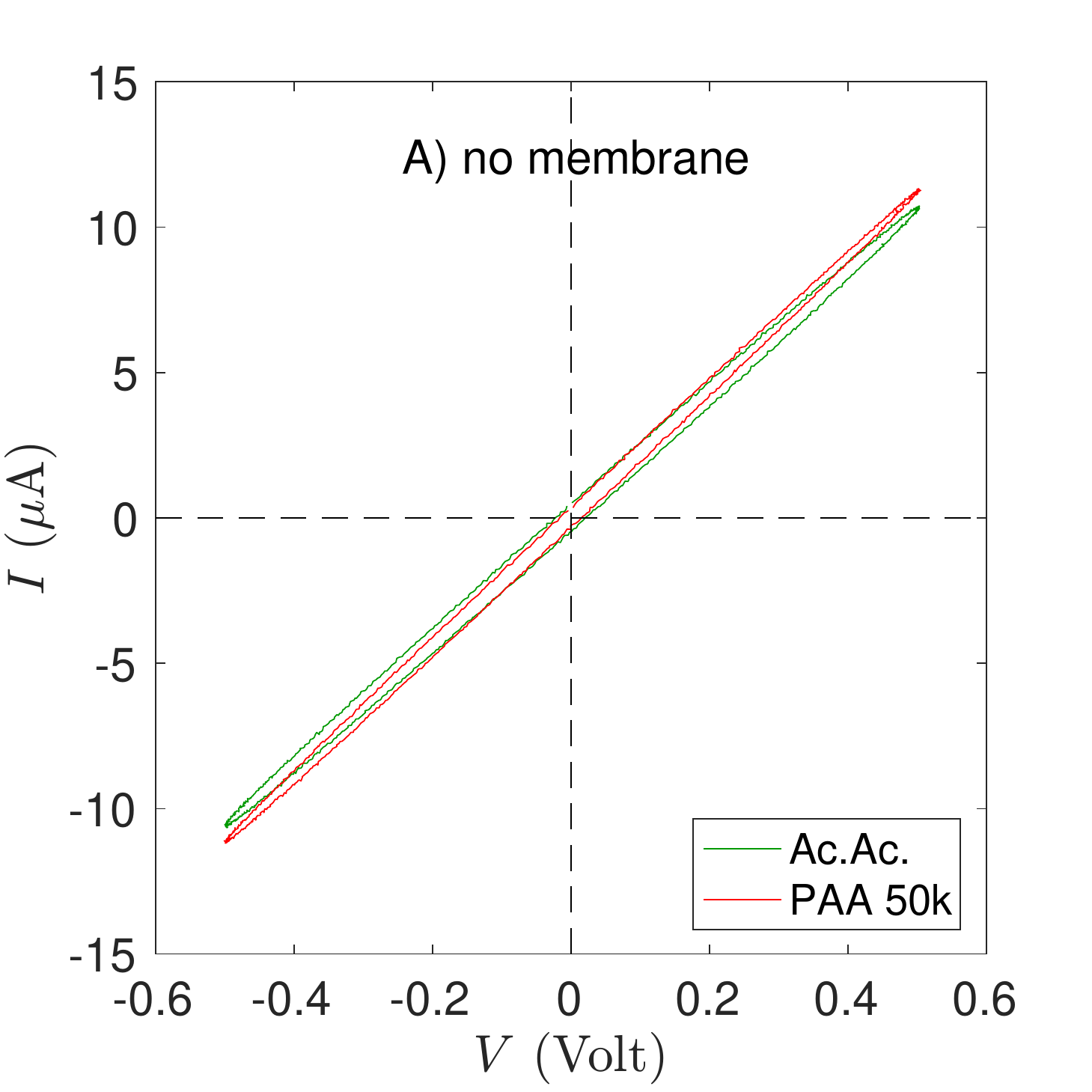}
\includegraphics[width=0.48\linewidth]{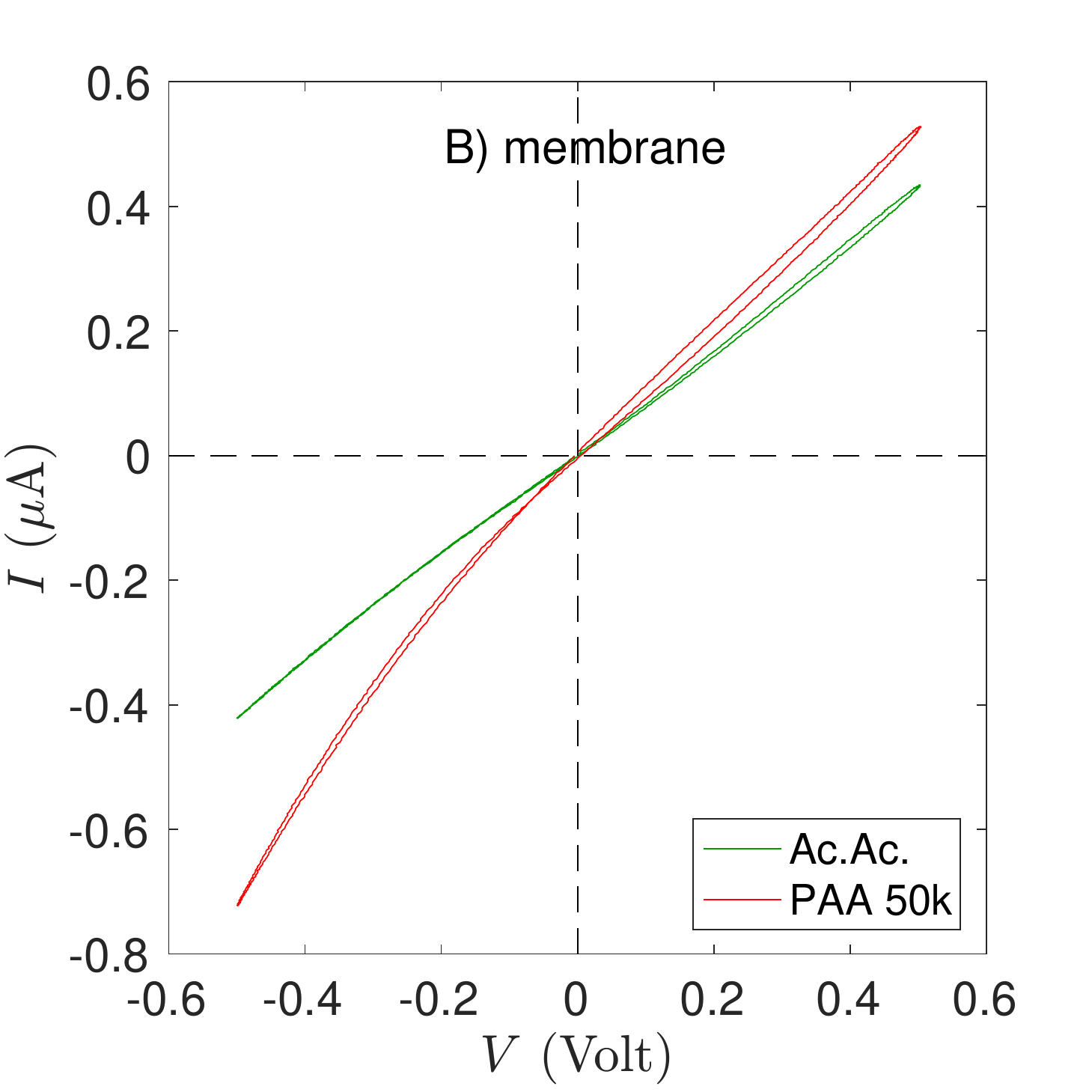}
\caption{Cyclic voltammetry at 1\,Hz for two electrolytes at pH=2.8$\pm$0.1: acetic acid at concentration $c=0.1$\,M (Ac.Ac.); polyacrylic acid, M=50$\times10^3$g/mol (PAA 50k) at monomer concentration $c=0.1$\,M. A)~no membrane between electrodes; B)~membrane with conical nanopores between electrodes.}
\label{fig2}
\end{figure}

\emph{Results and discussion: } Dealing with entropic effects on ions transport, the major issue comes from static electric charges on pore-wall: track-etching of polymer film produces ionized chemical groups at the surface of pore-wall. In case of polyimide, these groups are carboxylic acids and amines. Thus depending on pH, the net charge of pore-wall can be reversed and so the corresponding potential for free-ions\,\cite{Cervera:2006}. In Fig.\,\ref{figpH1}-A, the ionic current $I$ measured through polyimide conical nanopores is plotted as a function of the membrane voltage $V$ for small ions solutions at three different pH (KCl 0.1\,M in acetate buffer at pH=2 and 3.6, respectively, and acid acetic 0.1\,M pH=2.8). At pH=2, carboxylic acids and amines are protonated resulting in a positive net charge of pore-wall, while at pH=3.6 a sufficient quantity of carboxylic acids are dissociated to allow the inversion of the net charge. This is evidenced by a change of orientation of  ionic-current rectification (inversion of the concavity of $I=f(V)$). For acetic acid 0.1\,M (pH=2.8), no rectification is observed.
Rectification is better underlined by using the ``normalized rectification efficiency'' \cite{Kosinska:2008aa, Motz:2014aa}:
\begin{equation}\label{alpha}
\displaystyle\alpha(\|V\|)=\frac{i_--i_+}{i_-+i_+} 
\text{ with } \left\{{\begin{array}{l}
i_-=\|I(-\|V\|)\| \\ i_+=I(\|V\|)
\end{array}}\right.
\end{equation}
Note that $\alpha=0$ for a perfect resistor, whereas $\alpha=\pm1$ for a perfect diode depending on its direction. Results are plotted in Fig.\,\ref{figpH1}-B. At pH=2.8$\pm$0.1, rectification is almost negligible for small ions solutions. This pH is the isoelectric point of membranes. Results reported below were obtained at this pH, which is naturally obtained with polyacrylic acids ($\text{pK}_a\simeq4.5$\,\cite{Swift_2016} close to acetic acid) at a monomer concentration $c$=0.1\,M.

\begin{figure}[!htbp]
\centering
\includegraphics[width=0.8\linewidth]{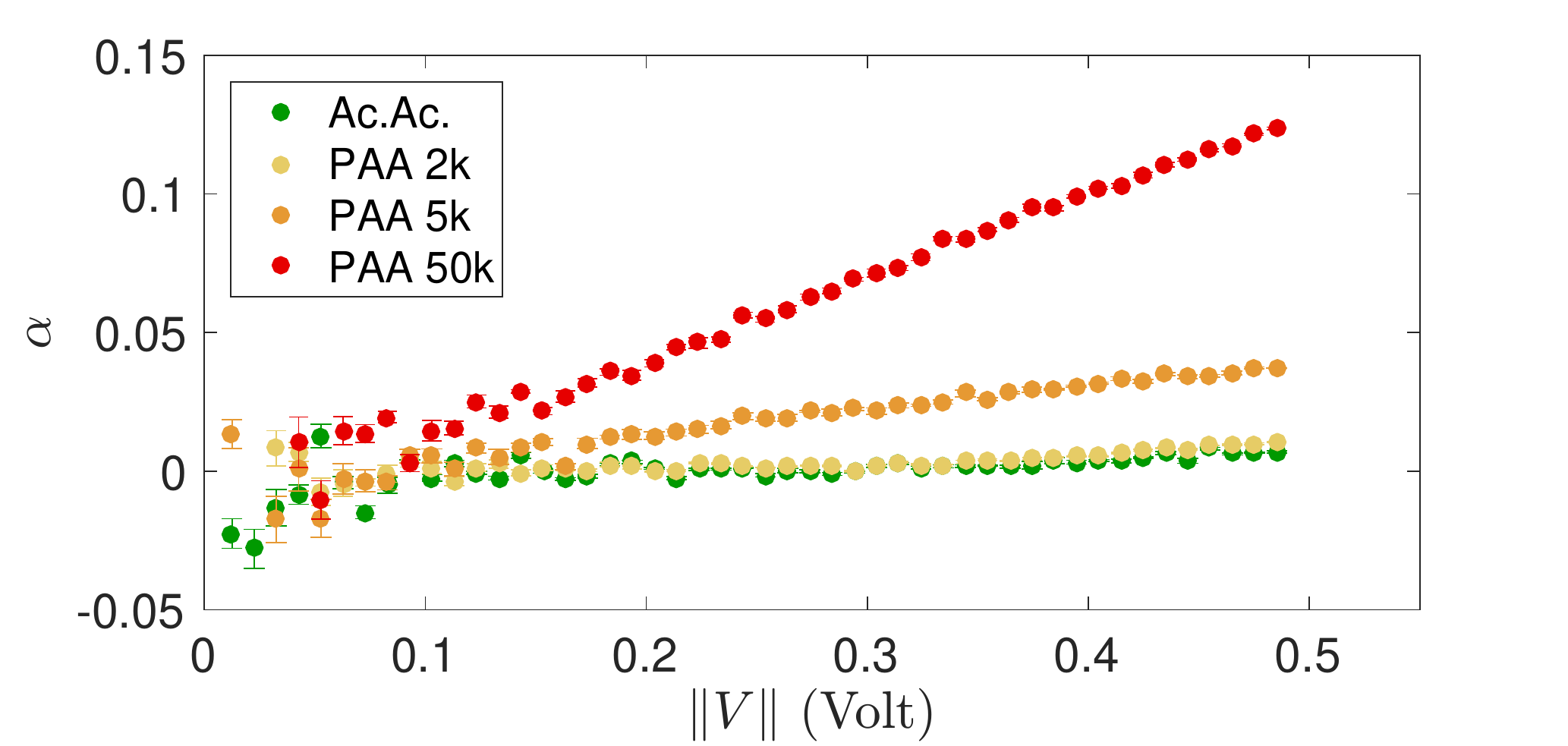}
\caption{Cyclic voltammetry at 1\,Hz: rectification efficiency $\alpha$ vs. absolute value of voltage $\|V\|$ (Eq.\,\ref{alpha}) for acetic acid (Ac.Ac., $c=0.1$\,M) and PAA ($M=2$, 5, and 50\,kg/mol, respectively; monomer concentration $c=0.1$\,M) measured with membrane at the isoelectric point  (pH=2.8$\pm$0.1).}
\label{alphavsU_PAA}
\end{figure}

Fig.\,\ref{fig2} shows cyclic voltammetry at 1\,Hz for polyacrylic acid (PAA) with molecular weight M=50$\times10^3$g/mol and acetic acid at the same concentration. Introducing a membrane with conical nanopores between electrodes, causes a current drop but also a patent rectification even though the membrane is at its isoelectric point. In Fig.\,\ref{alphavsU_PAA}, the rectification efficiency $\alpha$, is plotted as a function of $\|V\|$ for three different molecular weights $M$ and compared to acetic acid. The size dependence of the rectification efficiency is clear. This is our key result.

\begin{figure}[!htbp]
\centering
\includegraphics[width=0.8\linewidth]{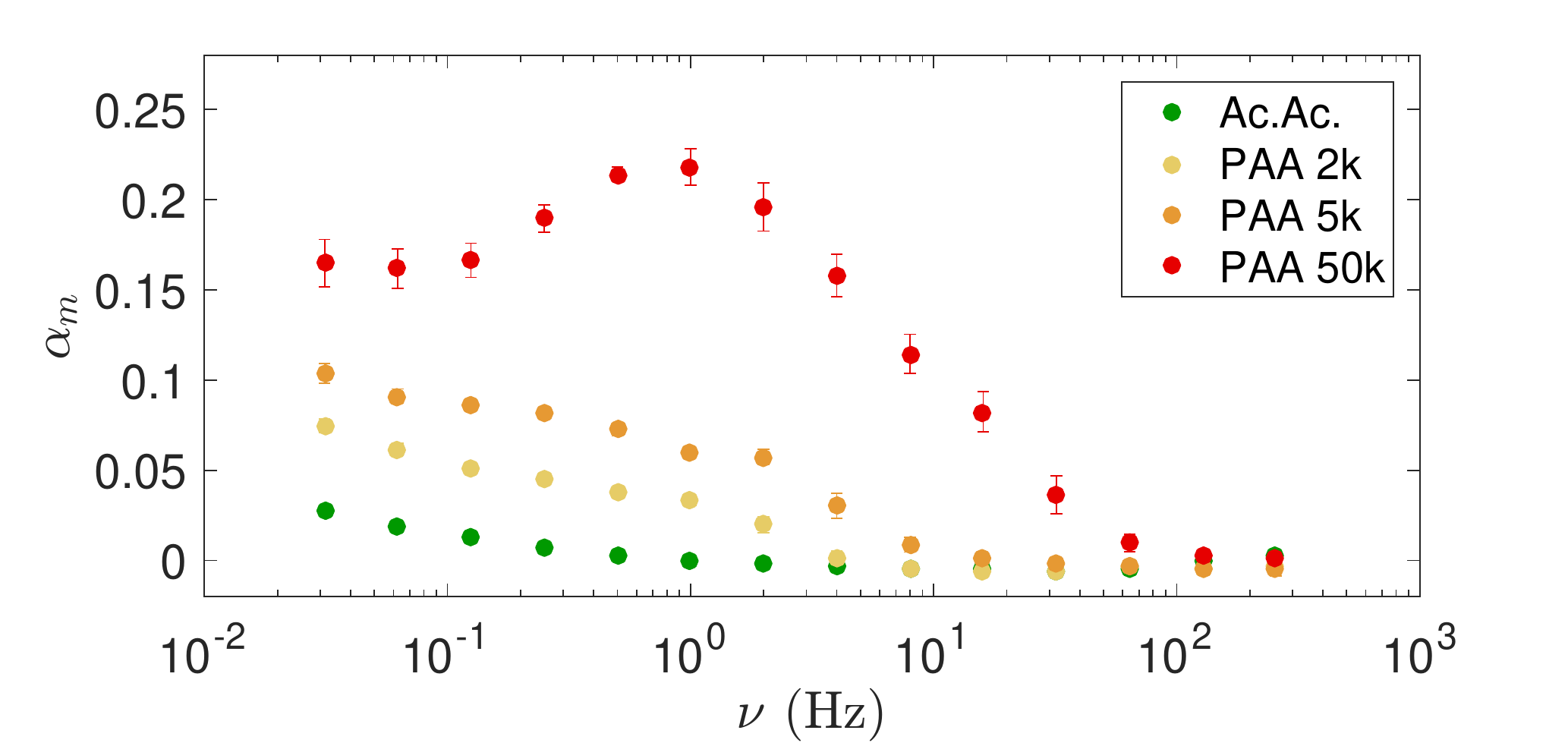}
\caption{Maximum rectification efficiency $\alpha_m$ (Eq.\,\ref{alpham}) vs. cycle frequency $\nu$ for acetic acid (Ac.Ac. $c=0.1$\,M) and PAA of different molecular weights ($M=2$, 5, and 50\,kg/mol, respectively; monomer concentration $c=0.1$\,M)  measured with membrane at the isoelectric point (pH=2.8$\pm$0.1).}
\label{alphavsF_PAA}
\end{figure}

Examination of $I=f(V)$ curves (see Fig.\,\ref{figpH1}--A and Fig.\ref{fig2}--B), shows that they tend to be linear for large enough $\|V\|$ values. Hence, $\alpha(\|V\|)$ tends to a maximum $\alpha_m$ that can be estimated by:
\begin{equation}\label{alpham}
\displaystyle\alpha_m=\frac{g_{-}-g_{+}}{g_{-}+g_{+}} \\ \\
\text{ with } \left\{{\begin{array}{l}
g_{-}=\left({dI/dV}\right)_{V<-0.4\text{V}} \\ g_{+}=\left({dI/dV}\right)_{V>0.4\text{V}}
\end{array}}\right.
\end{equation}
The ratio $\alpha_m$ is plotted versus cycle frequency in Fig.\,\ref{alphavsF_PAA} for acetic and PAA. We observe that  $\alpha_m$ globally decreases for increasing frequency, but also passes by a smooth maximum for the highest molecular weight anion. Also, spectra are shifted to higher frequency for increasing anion size.

\begin{figure}[!htbp]
\centering
\includegraphics[width=0.8\linewidth]{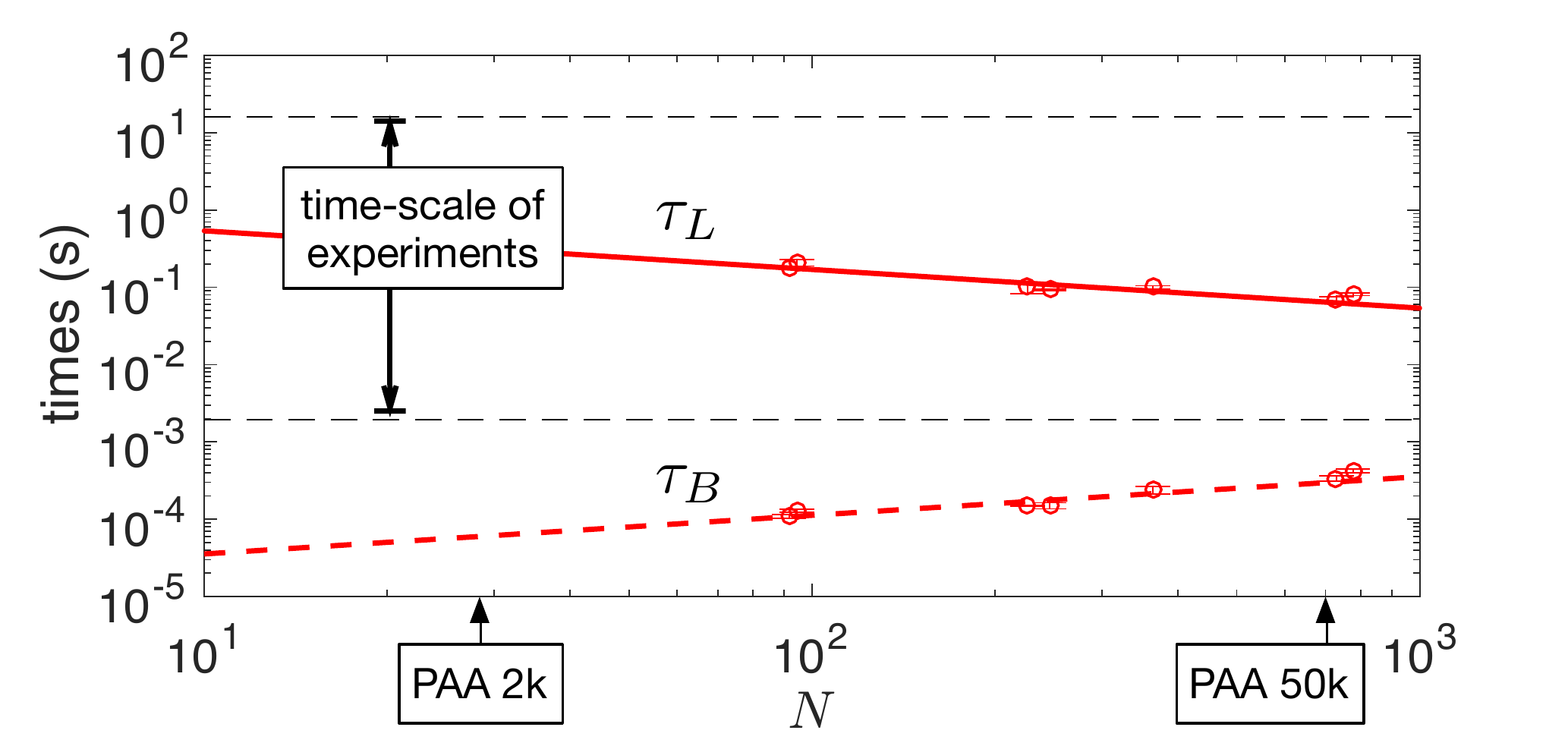}
\caption{Characteristic times of PAA in our conical nanopores vs. number $N$ of  segments per chain. $\tau_B$ (dashed red line) is the time needed for a chain to explore radially the pore (Eq.\,\ref{taub}). $\tau_L$ (full red line) is the time needed to cross the membrane for a voltage of 0.5\,V (Eq.\,\ref{taul}). Data points correspond to $\tau_B$  and $\tau_L$ calculated from data of ref.\,\cite{Swift_2016}. Horizontal dashed lines indicate the time-scale (frequency range) of  Fig.\,\ref{alphavsF_PAA}.}
\label{timesvsN}
\end{figure}

Let us focus on this last point and estimate the different characteristic times of our experimental system. 
The first to consider is the transverse equilibration time that can be identified as the one needed for one chain to diffuse over the large aperture $B$ of nanopore\,:
\begin{equation}\label{taub}
\tau_{B}=B^2/D
\end{equation}
This time has to be compared to the time $\tau_L$ needed to cross the channel. Due to the non-constant cross-section area $A(x)$, the electric field also is non-constant\,: $E(x)=V/L\times \pi Bb/A(x)$. For one chain experiencing this field the bias strength is $f=E(x)\gamma N\text{e}$, with $\text{e}$ the elementary charge, $\gamma$ the dissociation ratio of acid groups and $N$ the number of statistical segments per chain ($\propto$ molecular weight). In the stationary stage, this force is balanced by the friction $v(x) kT/D$, with $v(x)$ the chain velocity, hence $v(x)=\gamma N\text{e}E(x) D/kT$. Thus the time $\tau_L=\int_0^Lv^{-1}(x) dx$ to cross the entire pore writes\,:
\begin{equation}\label{taul}
\tau_L=\frac{kT}{V \gamma N\text{e} D}\times\frac{L}{\pi B b}\times\int_0^LA(x)dx
\end{equation}
From measurements reported in ref.\,\cite{Swift_2016}, one gets that at low pH the diffusion coefficient $D$ of PAA scales as $D=D_0\times N^{-1/2}$, with $D_0\simeq 10^{-9}$m$^2$/s. Note that $\tau_{B}$ increases for increasing chain length, whereas $\tau_L$ decreases. In Fig.\,\ref{timesvsN}, $\tau_B$ and $\tau_L$ were computed from experimental $D$ values\,\cite{Swift_2016} for $V=0.5$\,V, $L=6$\,$\mu$m, $\gamma\simeq2\times10^{-2}$ (i.e. pK$_a$=4.5 and $c=0.1$\,M), $B=100$\,nm and $b=2$\,nm and plotted as a function of $N$. 
First, one can see that in our experiments $\tau_B$ is always at least 100 times smaller than $\tau_L$. This gives ground for the 1D treatment at the base of Eq.\,\ref{FJeq}.
For cycle frequency $\nu$ higher than $\tau_L^{-1}$, polyanions do not have time to cross the entire nanopore and remain localized in a small slice. Hence, $\tau_L^{-1}$ would give a simple estimate of the cycle-frequency above which  rectification vanishes\,\cite{Motz:2014aa}. This may account for the shift of spectra to high frequency for increasing ion size (Fig.\,\ref{alphavsF_PAA}). Further attempt to rescale data of Fig.\,\ref{alphavsF_PAA} by using $\nu\tau_L$ as reduced variable is at risk, because we still lack the y-axis reduced variable that should be likely necessary. However, it is quite remarkable that the drop of rectification efficiency is found to lie in the time window calculated for $\tau_L$. This point plainly argues that the energy potential experienced by polyions extends over the full length $L$ of nanopores and consequently supports the idea of its entropic origin. Although the membrane is at the isoelectric point, let us consider the alternative origin that should be an electrostatic potential due to the pore-wall. Electrostatic interactions are screened above the Debye's length $\lambda_D\simeq7$\,nm for $c_{\text{ions}}=\gamma c=2\times 10^{-3}$\,M. So, the asymmetric electrostatic potential due to the pore-wall does not extend over the entire pore length but only over the length $l=\lambda_D\times L/(B-b)\simeq 400$\,nm from the cone tip. In Eq.\,\ref{taul}, the corresponding volume $\int_0^lA(x) dx$ is $1.5\times 10^3$ times smaller than the entire pore volume, and also the  time needed to cross this small region. This time is out of the time-window of our experiments and could not account for the shift of spectra reported in Fig.\,\ref{alphavsF_PAA}.

For the three PAA molecular weights here studied, the hydrodynamic radii $R_h$ 
are equal to 1.2, 1.9 and 6\,nm, respectivelly\,\cite{Swift_2016}.
As regards the pore size (small aperture radius $b\simeq2$\,nm), the question of chain blockade has to be examined. Actually, the electric field gradient due to the conical shape of pores is responsible for a stretching force applied on chains\,: $f_s=\gamma N\text{e}\times\nabla E(x)\times R$, with $R$ the radius of gyration of chains. This causes chains to elongate longitudinally but also to shrink laterally. In this latter direction, the radius of chain can be estimated as $r_\perp=R\times(kT/(f_s R/\sqrt N))^{1/6}$ (see \cite{deGennes:1996} part I.4). Taking $R\simeq R_h$, one gets the shrinking ratio $r_\perp/R_h\simeq0.3$ at the cone-tip (maximum field gradient) for the largest PAA chains, which should be sufficient to avoid chain blockades. Note that chain deformation does not weaken the previous calculation for $\tau_B$ and $\tau_L$ because deformation is only significant at the cone-tip and the time spent in this region is very short (the field is maximum) compared to the time spent in the rest of the nano-channel

Further analysis of our results in relation with theory is not straightforward (in particular the non-monotonic rectification spectrum in Fig.\,\ref{alphavsF_PAA}). Mainly because equation of motion through entropic barriers was not solved analytically, but with brownian dynamic simulations that amount to integrate Langevin equation in some particular cases that inevitably do not match to our experimental system\,\cite{Reguera:2006aa, Rubi:2010aa, Reguera:2012aa, Motz:2014aa} (e.g. a difference should be the coupling between the particle size and bias strength).
However, we have shown in this letter that rectification of ionic current can be obtained with electrically neutral conical nanopores in case anions and cations widely differ in size. Our observations depend on ions size and cycle frequency. 
Currently, the entropic ratchet mechanism is the only able to account for them.

We hope our results will stimulate researches in the field.
Also, we think that the notion of entropic ratchet is probably relevant for a better understanding of the facilitated or jammed-like transport observed for ionic-liquids confined in conical nanopores\,\cite{Tasserit:2010fk}, with likely outcomes for the improvement of electric batteries and cells that use confined geometries\,\cite{Berrod:2016aa}.


\begin{thebibliography}{29}%
\makeatletter
\providecommand \@ifxundefined [1]{%
 \@ifx{#1\undefined}
}%
\providecommand \@ifnum [1]{%
 \ifnum #1\expandafter \@firstoftwo
 \else \expandafter \@secondoftwo
 \fi
}%
\providecommand \@ifx [1]{%
 \ifx #1\expandafter \@firstoftwo
 \else \expandafter \@secondoftwo
 \fi
}%
\providecommand \natexlab [1]{#1}%
\providecommand \enquote  [1]{``#1''}%
\providecommand \bibnamefont  [1]{#1}%
\providecommand \bibfnamefont [1]{#1}%
\providecommand \citenamefont [1]{#1}%
\providecommand \href@noop [0]{\@secondoftwo}%
\providecommand \href [0]{\begingroup \@sanitize@url \@href}%
\providecommand \@href[1]{\@@startlink{#1}\@@href}%
\providecommand \@@href[1]{\endgroup#1\@@endlink}%
\providecommand \@sanitize@url [0]{\catcode `\\12\catcode `\$12\catcode
  `\&12\catcode `\#12\catcode `\^12\catcode `\_12\catcode `\%12\relax}%
\providecommand \@@startlink[1]{}%
\providecommand \@@endlink[0]{}%
\providecommand \url  [0]{\begingroup\@sanitize@url \@url }%
\providecommand \@url [1]{\endgroup\@href {#1}{\urlprefix }}%
\providecommand \urlprefix  [0]{URL }%
\providecommand \Eprint [0]{\href }%
\providecommand \doibase [0]{http://dx.doi.org/}%
\providecommand \selectlanguage [0]{\@gobble}%
\providecommand \bibinfo  [0]{\@secondoftwo}%
\providecommand \bibfield  [0]{\@secondoftwo}%
\providecommand \translation [1]{[#1]}%
\providecommand \BibitemOpen [0]{}%
\providecommand \bibitemStop [0]{}%
\providecommand \bibitemNoStop [0]{.\EOS\space}%
\providecommand \EOS [0]{\spacefactor3000\relax}%
\providecommand \BibitemShut  [1]{\csname bibitem#1\endcsname}%
\let\auto@bib@innerbib\@empty
\bibitem [{\citenamefont {Astumian}(1997)}]{Astumian:1997aa}%
  \BibitemOpen
  \bibfield  {author} {\bibinfo {author} {\bibfnamefont {R.~D.}\ \bibnamefont
  {Astumian}},\ }\href
  {http://science.sciencemag.org/content/276/5314/917.abstract} {\bibfield
  {journal} {\bibinfo  {journal} {Science}\ }\textbf {\bibinfo {volume}
  {276}},\ \bibinfo {pages} {917} (\bibinfo {year} {1997})}\BibitemShut
  {NoStop}%
\bibitem [{\citenamefont {Reimann}(2002)}]{Reimann:2002aa}%
  \BibitemOpen
  \bibfield  {author} {\bibinfo {author} {\bibfnamefont {P.}~\bibnamefont
  {Reimann}},\ }\href {\doibase
  http://dx.doi.org/10.1016/S0370-1573(01)00081-3} {\bibfield  {journal}
  {\bibinfo  {journal} {Physics Reports}\ }\textbf {\bibinfo {volume} {361}},\
  \bibinfo {pages} {57} (\bibinfo {year} {2002})}\BibitemShut {NoStop}%
\bibitem [{\citenamefont {H{\"a}nggi}\ and\ \citenamefont
  {Marchesoni}(2009)}]{Hanggi:2009kx}%
  \BibitemOpen
  \bibfield  {author} {\bibinfo {author} {\bibfnamefont {P.}~\bibnamefont
  {H{\"a}nggi}}\ and\ \bibinfo {author} {\bibfnamefont {F.}~\bibnamefont
  {Marchesoni}},\ }\href {http://link.aps.org/doi/10.1103/RevModPhys.81.387}
  {\bibfield  {journal} {\bibinfo  {journal} {Rev. Mod. Phys.}\ }\textbf
  {\bibinfo {volume} {81}} (\bibinfo {year} {2009})}\BibitemShut {NoStop}%
\bibitem [{\citenamefont {Bridgman}(1928)}]{Bridgman:1928aa}%
  \BibitemOpen
  \bibfield  {author} {\bibinfo {author} {\bibfnamefont {P.~W.}\ \bibnamefont
  {Bridgman}},\ }\href {http://link.aps.org/doi/10.1103/PhysRev.31.101}
  {\bibfield  {journal} {\bibinfo  {journal} {Phys. Rev.}\ }\textbf {\bibinfo
  {volume} {31}},\ \bibinfo {pages} {101} (\bibinfo {year} {1928})}\BibitemShut
  {NoStop}%
\bibitem [{\citenamefont {Brillouin}(1950)}]{Brillouin:1950aa}%
  \BibitemOpen
  \bibfield  {author} {\bibinfo {author} {\bibfnamefont {L.}~\bibnamefont
  {Brillouin}},\ }\href {http://link.aps.org/doi/10.1103/PhysRev.78.627.2}
  {\bibfield  {journal} {\bibinfo  {journal} {Phys. Rev.}\ }\textbf {\bibinfo
  {volume} {78}},\ \bibinfo {pages} {627} (\bibinfo {year} {1950})}\BibitemShut
  {NoStop}%
\bibitem [{\citenamefont {Feynman}\ \emph {et~al.}(1966)\citenamefont
  {Feynman}, \citenamefont {Leighton},\ and\ \citenamefont
  {Sands}}]{Feynmann_Ratchet}%
  \BibitemOpen
  \bibfield  {author} {\bibinfo {author} {\bibfnamefont {R.~P.}\ \bibnamefont
  {Feynman}}, \bibinfo {author} {\bibfnamefont {R.~B.}\ \bibnamefont
  {Leighton}}, \ and\ \bibinfo {author} {\bibfnamefont {M.}~\bibnamefont
  {Sands}},\ }\enquote {\bibinfo {title} {The {F}eynman lectures on physics :
  Ratchet and pawl},}\ \ (\bibinfo  {publisher} {Addison-Wesley, Reading, MA},\
  \bibinfo {year} {1966})\ Chap.~\bibinfo {chapter} {46}\BibitemShut {NoStop}%
\bibitem [{\citenamefont {Vale}\ and\ \citenamefont
  {Oosawa}(1990)}]{Vale:1990aa}%
  \BibitemOpen
  \bibfield  {author} {\bibinfo {author} {\bibfnamefont {R.~D.}\ \bibnamefont
  {Vale}}\ and\ \bibinfo {author} {\bibfnamefont {F.}~\bibnamefont {Oosawa}},\
  }\href {\doibase http://dx.doi.org/} {\bibfield  {journal} {\bibinfo
  {journal} {Adv. Biophys.}\ }\textbf {\bibinfo {volume} {26}},\ \bibinfo
  {pages} {97} (\bibinfo {year} {1990})}\BibitemShut {NoStop}%
\bibitem [{\citenamefont {Astumian}\ and\ \citenamefont
  {Bier}(1994)}]{Astumian:1994aa}%
  \BibitemOpen
  \bibfield  {author} {\bibinfo {author} {\bibfnamefont {R.~D.}\ \bibnamefont
  {Astumian}}\ and\ \bibinfo {author} {\bibfnamefont {M.}~\bibnamefont
  {Bier}},\ }\href {http://link.aps.org/doi/10.1103/PhysRevLett.72.1766}
  {\bibfield  {journal} {\bibinfo  {journal} {Phys. Rev. Lett.}\ }\textbf
  {\bibinfo {volume} {72}},\ \bibinfo {pages} {1766} (\bibinfo {year}
  {1994})}\BibitemShut {NoStop}%
\bibitem [{\citenamefont {Tsong}\ and\ \citenamefont
  {Xie}(2002)}]{Tsong:2002aa}%
  \BibitemOpen
  \bibfield  {author} {\bibinfo {author} {\bibfnamefont {T.~Y.}\ \bibnamefont
  {Tsong}}\ and\ \bibinfo {author} {\bibfnamefont {T.~D.}\ \bibnamefont
  {Xie}},\ }\href {\doibase 10.1007/s003390201407} {\bibfield  {journal}
  {\bibinfo  {journal} {Appl. Phys. A}\ }\textbf {\bibinfo {volume} {75}},\
  \bibinfo {pages} {345} (\bibinfo {year} {2002})}\BibitemShut {NoStop}%
\bibitem [{\citenamefont {Fadda}\ \emph {et~al.}(2013)\citenamefont {Fadda},
  \citenamefont {Lairez}, \citenamefont {Guennouni},\ and\ \citenamefont
  {Koutsioubas}}]{Fadda:2013fk}%
  \BibitemOpen
  \bibfield  {author} {\bibinfo {author} {\bibfnamefont {G.~C.}\ \bibnamefont
  {Fadda}}, \bibinfo {author} {\bibfnamefont {D.}~\bibnamefont {Lairez}},
  \bibinfo {author} {\bibfnamefont {Z.}~\bibnamefont {Guennouni}}, \ and\
  \bibinfo {author} {\bibfnamefont {A.}~\bibnamefont {Koutsioubas}},\ }\href
  {http://link.aps.org/doi/10.1103/PhysRevLett.111.028102} {\bibfield
  {journal} {\bibinfo  {journal} {Phys. Rev. Lett.}\ }\textbf {\bibinfo
  {volume} {111}},\ \bibinfo {pages} {028102} (\bibinfo {year}
  {2013})}\BibitemShut {NoStop}%
\bibitem [{\citenamefont {Hou}\ \emph {et~al.}(2012)\citenamefont {Hou},
  \citenamefont {Zhang},\ and\ \citenamefont {Jiang}}]{Hou:2012aa}%
  \BibitemOpen
  \bibfield  {author} {\bibinfo {author} {\bibfnamefont {X.}~\bibnamefont
  {Hou}}, \bibinfo {author} {\bibfnamefont {H.}~\bibnamefont {Zhang}}, \ and\
  \bibinfo {author} {\bibfnamefont {L.}~\bibnamefont {Jiang}},\ }\href
  {\doibase 10.1002/anie.201104904} {\bibfield  {journal} {\bibinfo  {journal}
  {Angew. Chem. Int. Ed.}\ }\textbf {\bibinfo {volume} {51}},\ \bibinfo {pages}
  {5296} (\bibinfo {year} {2012})}\BibitemShut {NoStop}%
\bibitem [{\citenamefont {Siwy}\ and\ \citenamefont
  {Fuli\'nski}(2002)}]{Siwy:2002fk}%
  \BibitemOpen
  \bibfield  {author} {\bibinfo {author} {\bibfnamefont {Z.}~\bibnamefont
  {Siwy}}\ and\ \bibinfo {author} {\bibfnamefont {A.}~\bibnamefont
  {Fuli\'nski}},\ }\href@noop {} {\bibfield  {journal} {\bibinfo  {journal}
  {Phys. Rev. Lett.}\ }\textbf {\bibinfo {volume} {89}},\ \bibinfo {pages}
  {198103} (\bibinfo {year} {2002})}\BibitemShut {NoStop}%
\bibitem [{\citenamefont {Siwy}(2006)}]{Siwy:2006}%
  \BibitemOpen
  \bibfield  {author} {\bibinfo {author} {\bibfnamefont {Z.}~\bibnamefont
  {Siwy}},\ }\href@noop {} {\bibfield  {journal} {\bibinfo  {journal} {Adv.
  Funct. Mat.}\ }\textbf {\bibinfo {volume} {16}},\ \bibinfo {pages} {735}
  (\bibinfo {year} {2006})}\BibitemShut {NoStop}%
\bibitem [{\citenamefont {Cervera}\ \emph {et~al.}(2006)\citenamefont
  {Cervera}, \citenamefont {Schiedt}, \citenamefont {Neumann}, \citenamefont
  {Maf{\'e}},\ and\ \citenamefont {Ram\'irez}}]{Cervera:2006}%
  \BibitemOpen
  \bibfield  {author} {\bibinfo {author} {\bibfnamefont {J.}~\bibnamefont
  {Cervera}}, \bibinfo {author} {\bibfnamefont {B.}~\bibnamefont {Schiedt}},
  \bibinfo {author} {\bibfnamefont {R.}~\bibnamefont {Neumann}}, \bibinfo
  {author} {\bibfnamefont {S.}~\bibnamefont {Maf{\'e}}}, \ and\ \bibinfo
  {author} {\bibfnamefont {P.}~\bibnamefont {Ram\'irez}},\ }\href@noop {}
  {\bibfield  {journal} {\bibinfo  {journal} {J. Chem. Phys.}\ }\textbf
  {\bibinfo {volume} {124}},\ \bibinfo {pages} {104706} (\bibinfo {year}
  {2006})}\BibitemShut {NoStop}%
\bibitem [{\citenamefont {Sheng}\ \emph {et~al.}(2014)\citenamefont {Sheng},
  \citenamefont {Wang}, \citenamefont {Wang}, \citenamefont {Wang},\ and\
  \citenamefont {Xue}}]{sheng2014}%
  \BibitemOpen
  \bibfield  {author} {\bibinfo {author} {\bibfnamefont {Q.}~\bibnamefont
  {Sheng}}, \bibinfo {author} {\bibfnamefont {L.}~\bibnamefont {Wang}},
  \bibinfo {author} {\bibfnamefont {C.}~\bibnamefont {Wang}}, \bibinfo {author}
  {\bibfnamefont {X.}~\bibnamefont {Wang}}, \ and\ \bibinfo {author}
  {\bibfnamefont {J.}~\bibnamefont {Xue}},\ }\href
  {http://scitation.aip.org/content/aip/journal/bmf/8/5/10.1063/1.4896474}
  {\bibfield  {journal} {\bibinfo  {journal} {Biomicrofluidics}\ }\textbf
  {\bibinfo {volume} {8}} (\bibinfo {year} {2014})}\BibitemShut {NoStop}%
\bibitem [{\citenamefont {Zhang}\ \emph {et~al.}(2015)\citenamefont {Zhang},
  \citenamefont {Kong}, \citenamefont {Xiao}, \citenamefont {Liu},
  \citenamefont {Xie}, \citenamefont {Li}, \citenamefont {Ma}, \citenamefont
  {Tian}, \citenamefont {Wen},\ and\ \citenamefont {Jiang}}]{Zhang:2015aa}%
  \BibitemOpen
  \bibfield  {author} {\bibinfo {author} {\bibfnamefont {Z.}~\bibnamefont
  {Zhang}}, \bibinfo {author} {\bibfnamefont {X.-Y.}\ \bibnamefont {Kong}},
  \bibinfo {author} {\bibfnamefont {K.}~\bibnamefont {Xiao}}, \bibinfo {author}
  {\bibfnamefont {Q.}~\bibnamefont {Liu}}, \bibinfo {author} {\bibfnamefont
  {G.}~\bibnamefont {Xie}}, \bibinfo {author} {\bibfnamefont {P.}~\bibnamefont
  {Li}}, \bibinfo {author} {\bibfnamefont {J.}~\bibnamefont {Ma}}, \bibinfo
  {author} {\bibfnamefont {Y.}~\bibnamefont {Tian}}, \bibinfo {author}
  {\bibfnamefont {L.}~\bibnamefont {Wen}}, \ and\ \bibinfo {author}
  {\bibfnamefont {L.}~\bibnamefont {Jiang}},\ }\href {\doibase
  10.1021/jacs.5b09918} {\bibfield  {journal} {\bibinfo  {journal} {J. Am.
  Chem. Soc.}\ }\textbf {\bibinfo {volume} {137}},\ \bibinfo {pages} {14765}
  (\bibinfo {year} {2015})}\BibitemShut {NoStop}%
\bibitem [{\citenamefont {Madrid}\ \emph {et~al.}(2015)\citenamefont {Madrid},
  \citenamefont {Cottis}, \citenamefont {Rong}, \citenamefont {Rogers},
  \citenamefont {Stone}, \citenamefont {Malpass-Evans}, \citenamefont {Carta},
  \citenamefont {McKeown},\ and\ \citenamefont {Marken}}]{Madrid:2015aa}%
  \BibitemOpen
  \bibfield  {author} {\bibinfo {author} {\bibfnamefont {E.}~\bibnamefont
  {Madrid}}, \bibinfo {author} {\bibfnamefont {P.}~\bibnamefont {Cottis}},
  \bibinfo {author} {\bibfnamefont {Y.}~\bibnamefont {Rong}}, \bibinfo {author}
  {\bibfnamefont {A.~T.}\ \bibnamefont {Rogers}}, \bibinfo {author}
  {\bibfnamefont {J.~M.}\ \bibnamefont {Stone}}, \bibinfo {author}
  {\bibfnamefont {R.}~\bibnamefont {Malpass-Evans}}, \bibinfo {author}
  {\bibfnamefont {M.}~\bibnamefont {Carta}}, \bibinfo {author} {\bibfnamefont
  {N.~B.}\ \bibnamefont {McKeown}}, \ and\ \bibinfo {author} {\bibfnamefont
  {F.}~\bibnamefont {Marken}},\ }\href {\doibase 10.1039/C5TA04092B} {\bibfield
   {journal} {\bibinfo  {journal} {J. Mat. Chem. A}\ }\textbf {\bibinfo
  {volume} {3}},\ \bibinfo {pages} {15849} (\bibinfo {year}
  {2015})}\BibitemShut {NoStop}%
\bibitem [{\citenamefont {Jiang}\ \emph {et~al.}(2016)\citenamefont {Jiang},
  \citenamefont {Liu},\ and\ \citenamefont {Qiao}}]{Jiang:2016aa}%
  \BibitemOpen
  \bibfield  {author} {\bibinfo {author} {\bibfnamefont {X.}~\bibnamefont
  {Jiang}}, \bibinfo {author} {\bibfnamefont {Y.}~\bibnamefont {Liu}}, \ and\
  \bibinfo {author} {\bibfnamefont {R.}~\bibnamefont {Qiao}},\ }\href {\doibase
  10.1021/acs.jpcc.5b11522} {\bibfield  {journal} {\bibinfo  {journal} {J.
  Phys. Chem. C}\ }\textbf {\bibinfo {volume} {120}},\ \bibinfo {pages} {4629}
  (\bibinfo {year} {2016})}\BibitemShut {NoStop}%
\bibitem [{\citenamefont {Kosi{\'n}ska}\ \emph {et~al.}(2008)\citenamefont
  {Kosi{\'n}ska}, \citenamefont {Goychuk}, \citenamefont {Kostur},
  \citenamefont {Schmid},\ and\ \citenamefont {H{\"a}nggi}}]{Kosinska:2008aa}%
  \BibitemOpen
  \bibfield  {author} {\bibinfo {author} {\bibfnamefont {I.~D.}\ \bibnamefont
  {Kosi{\'n}ska}}, \bibinfo {author} {\bibfnamefont {I.}~\bibnamefont
  {Goychuk}}, \bibinfo {author} {\bibfnamefont {M.}~\bibnamefont {Kostur}},
  \bibinfo {author} {\bibfnamefont {G.}~\bibnamefont {Schmid}}, \ and\ \bibinfo
  {author} {\bibfnamefont {P.}~\bibnamefont {H{\"a}nggi}},\ }\href
  {http://link.aps.org/doi/10.1103/PhysRevE.77.031131} {\bibfield  {journal}
  {\bibinfo  {journal} {Phys. Rev. E}\ }\textbf {\bibinfo {volume} {77}},\
  \bibinfo {pages} {031131} (\bibinfo {year} {2008})}\BibitemShut {NoStop}%
\bibitem [{\citenamefont {Reguera}\ \emph {et~al.}(2006)\citenamefont
  {Reguera}, \citenamefont {Schmid}, \citenamefont {Burada}, \citenamefont
  {Rub\'{i}}, \citenamefont {Reimann},\ and\ \citenamefont
  {H\"{a}nggi}}]{Reguera:2006aa}%
  \BibitemOpen
  \bibfield  {author} {\bibinfo {author} {\bibfnamefont {D.}~\bibnamefont
  {Reguera}}, \bibinfo {author} {\bibfnamefont {G.}~\bibnamefont {Schmid}},
  \bibinfo {author} {\bibfnamefont {P.~S.}\ \bibnamefont {Burada}}, \bibinfo
  {author} {\bibfnamefont {J.~M.}\ \bibnamefont {Rub\'{i}}}, \bibinfo {author}
  {\bibfnamefont {P.}~\bibnamefont {Reimann}}, \ and\ \bibinfo {author}
  {\bibfnamefont {P.}~\bibnamefont {H\"{a}nggi}},\ }\href
  {http://link.aps.org/doi/10.1103/PhysRevLett.96.130603} {\bibfield  {journal}
  {\bibinfo  {journal} {Phys. Rev. Lett.}\ }\textbf {\bibinfo {volume} {96}},\
  \bibinfo {pages} {130603} (\bibinfo {year} {2006})}\BibitemShut {NoStop}%
\bibitem [{\citenamefont {de~Gennes}(1996)}]{deGennes:1996}%
  \BibitemOpen
  \bibfield  {author} {\bibinfo {author} {\bibfnamefont {P.~G.}\ \bibnamefont
  {de~Gennes}},\ }\href@noop {} {\emph {\bibinfo {title} {Scaling concepts in
  polymer physics.}}}\ (\bibinfo  {publisher} {Cornell Univ. Press},\ \bibinfo
  {address} {Ithaca},\ \bibinfo {year} {1996})\BibitemShut {NoStop}%
\bibitem [{\citenamefont {Mao}\ \emph {et~al.}(1995)\citenamefont {Mao},
  \citenamefont {Cates},\ and\ \citenamefont {Lekkerkerker}}]{Mao:1995aa}%
  \BibitemOpen
  \bibfield  {author} {\bibinfo {author} {\bibfnamefont {Y.}~\bibnamefont
  {Mao}}, \bibinfo {author} {\bibfnamefont {M.~E.}\ \bibnamefont {Cates}}, \
  and\ \bibinfo {author} {\bibfnamefont {H.~N.~W.}\ \bibnamefont
  {Lekkerkerker}},\ }\href {\doibase
  http://dx.doi.org/10.1016/0378-4371(95)00206-5} {\bibfield  {journal}
  {\bibinfo  {journal} {Physica A}\ }\textbf {\bibinfo {volume} {222}},\
  \bibinfo {pages} {10} (\bibinfo {year} {1995})}\BibitemShut {NoStop}%
\bibitem [{\citenamefont {Zwanzig}(1992)}]{Zwanzig:1992aa}%
  \BibitemOpen
  \bibfield  {author} {\bibinfo {author} {\bibfnamefont {R.}~\bibnamefont
  {Zwanzig}},\ }\href {\doibase 10.1021/j100189a004} {\bibfield  {journal}
  {\bibinfo  {journal} {J. Phys. Chem.}\ }\textbf {\bibinfo {volume} {96}},\
  \bibinfo {pages} {3926} (\bibinfo {year} {1992})}\BibitemShut {NoStop}%
\bibitem [{\citenamefont {Rub\'{i}}\ and\ \citenamefont
  {Reguera}(2010)}]{Rubi:2010aa}%
  \BibitemOpen
  \bibfield  {author} {\bibinfo {author} {\bibfnamefont {J.~M.}\ \bibnamefont
  {Rub\'{i}}}\ and\ \bibinfo {author} {\bibfnamefont {D.}~\bibnamefont
  {Reguera}},\ }\href {\doibase
  http://dx.doi.org/10.1016/j.chemphys.2010.04.029} {\bibfield  {journal}
  {\bibinfo  {journal} {Chem. Phys.}\ }\textbf {\bibinfo {volume} {375}},\
  \bibinfo {pages} {518} (\bibinfo {year} {2010})}\BibitemShut {NoStop}%
\bibitem [{\citenamefont {Reguera}\ \emph {et~al.}(2012)\citenamefont
  {Reguera}, \citenamefont {Luque}, \citenamefont {Burada}, \citenamefont
  {Schmid}, \citenamefont {Rub\'{i}},\ and\ \citenamefont
  {H\"{a}nggi}}]{Reguera:2012aa}%
  \BibitemOpen
  \bibfield  {author} {\bibinfo {author} {\bibfnamefont {D.}~\bibnamefont
  {Reguera}}, \bibinfo {author} {\bibfnamefont {A.}~\bibnamefont {Luque}},
  \bibinfo {author} {\bibfnamefont {P.~S.}\ \bibnamefont {Burada}}, \bibinfo
  {author} {\bibfnamefont {G.}~\bibnamefont {Schmid}}, \bibinfo {author}
  {\bibfnamefont {J.~M.}\ \bibnamefont {Rub\'{i}}}, \ and\ \bibinfo {author}
  {\bibfnamefont {P.}~\bibnamefont {H\"{a}nggi}},\ }\href
  {http://link.aps.org/doi/10.1103/PhysRevLett.108.020604} {\bibfield
  {journal} {\bibinfo  {journal} {Phys. Rev. Lett.}\ }\textbf {\bibinfo
  {volume} {108}},\ \bibinfo {pages} {020604} (\bibinfo {year}
  {2012})}\BibitemShut {NoStop}%
\bibitem [{\citenamefont {Tasserit}\ \emph {et~al.}(2010)\citenamefont
  {Tasserit}, \citenamefont {Koutsioubas}, \citenamefont {Lairez},
  \citenamefont {Zalczer},\ and\ \citenamefont {Clochard}}]{Tasserit:2010fk}%
  \BibitemOpen
  \bibfield  {author} {\bibinfo {author} {\bibfnamefont {C.}~\bibnamefont
  {Tasserit}}, \bibinfo {author} {\bibfnamefont {A.}~\bibnamefont
  {Koutsioubas}}, \bibinfo {author} {\bibfnamefont {D.}~\bibnamefont {Lairez}},
  \bibinfo {author} {\bibfnamefont {G.}~\bibnamefont {Zalczer}}, \ and\
  \bibinfo {author} {\bibfnamefont {M.-C.}\ \bibnamefont {Clochard}},\
  }\href@noop {} {\bibfield  {journal} {\bibinfo  {journal} {Phys. Rev. Lett.}\
  }\textbf {\bibinfo {volume} {105}},\ \bibinfo {pages} {260602} (\bibinfo
  {year} {2010})}\BibitemShut {NoStop}%
\bibitem [{\citenamefont {Motz}\ \emph {et~al.}(2014)\citenamefont {Motz},
  \citenamefont {Schmid}, \citenamefont {H{\"a}nggi}, \citenamefont {Reguera},\
  and\ \citenamefont {Rub{\'\i}}}]{Motz:2014aa}%
  \BibitemOpen
  \bibfield  {author} {\bibinfo {author} {\bibfnamefont {T.}~\bibnamefont
  {Motz}}, \bibinfo {author} {\bibfnamefont {G.}~\bibnamefont {Schmid}},
  \bibinfo {author} {\bibfnamefont {P.}~\bibnamefont {H{\"a}nggi}}, \bibinfo
  {author} {\bibfnamefont {D.}~\bibnamefont {Reguera}}, \ and\ \bibinfo
  {author} {\bibfnamefont {J.~M.}\ \bibnamefont {Rub{\'\i}}},\ }\href
  {http://scitation.aip.org/content/aip/journal/jcp/141/7/10.1063/1.4892615}
  {\bibfield  {journal} {\bibinfo  {journal} {J. Chem. Phys.}\ }\textbf
  {\bibinfo {volume} {141}},\ \bibinfo {pages} {074104} (\bibinfo {year}
  {2014})}\BibitemShut {NoStop}%
\bibitem [{\citenamefont {Swift}\ \emph {et~al.}(2016)\citenamefont {Swift},
  \citenamefont {Swanson}, \citenamefont {Geoghegan},\ and\ \citenamefont
  {Rimmer}}]{Swift_2016}%
  \BibitemOpen
  \bibfield  {author} {\bibinfo {author} {\bibfnamefont {T.}~\bibnamefont
  {Swift}}, \bibinfo {author} {\bibfnamefont {L.}~\bibnamefont {Swanson}},
  \bibinfo {author} {\bibfnamefont {M.}~\bibnamefont {Geoghegan}}, \ and\
  \bibinfo {author} {\bibfnamefont {S.}~\bibnamefont {Rimmer}},\ }\href
  {\doibase 10.1039/C5SM02693H} {\bibfield  {journal} {\bibinfo  {journal}
  {Soft Matter}\ }\textbf {\bibinfo {volume} {12}},\ \bibinfo {pages} {2542}
  (\bibinfo {year} {2016})}\BibitemShut {NoStop}%
\bibitem [{\citenamefont {Berrod}\ \emph {et~al.}(2016)\citenamefont {Berrod},
  \citenamefont {Ferdeghini}, \citenamefont {Judeinstein}, \citenamefont
  {Genevaz}, \citenamefont {Ramos}, \citenamefont {Fournier}, \citenamefont
  {Dijon}, \citenamefont {Ollivier}, \citenamefont {Rols}, \citenamefont {Yu},
  \citenamefont {Mole},\ and\ \citenamefont {Zanotti}}]{Berrod:2016aa}%
  \BibitemOpen
  \bibfield  {author} {\bibinfo {author} {\bibfnamefont {Q.}~\bibnamefont
  {Berrod}}, \bibinfo {author} {\bibfnamefont {F.}~\bibnamefont {Ferdeghini}},
  \bibinfo {author} {\bibfnamefont {P.}~\bibnamefont {Judeinstein}}, \bibinfo
  {author} {\bibfnamefont {N.}~\bibnamefont {Genevaz}}, \bibinfo {author}
  {\bibfnamefont {R.}~\bibnamefont {Ramos}}, \bibinfo {author} {\bibfnamefont
  {A.}~\bibnamefont {Fournier}}, \bibinfo {author} {\bibfnamefont
  {J.}~\bibnamefont {Dijon}}, \bibinfo {author} {\bibfnamefont
  {J.}~\bibnamefont {Ollivier}}, \bibinfo {author} {\bibfnamefont
  {S.}~\bibnamefont {Rols}}, \bibinfo {author} {\bibfnamefont {D.}~\bibnamefont
  {Yu}}, \bibinfo {author} {\bibfnamefont {R.~A.}\ \bibnamefont {Mole}}, \ and\
  \bibinfo {author} {\bibfnamefont {J.~M.}\ \bibnamefont {Zanotti}},\ }\href
  {\doibase 10.1039/C6NR01445C} {\bibfield  {journal} {\bibinfo  {journal}
  {Nanoscale}\ }\textbf {\bibinfo {volume} {8}},\ \bibinfo {pages} {7845}
  (\bibinfo {year} {2016})}\BibitemShut {NoStop}%
\end{thebibliography}
%

\end{document}